\documentstyle{article}
\begin{document}
\title{Decoherence and quantum mechanics in a gravitational field}
\author{Z. Haba\\Institute of Theoretical Physics,University of
Wroclaw,Wroclaw,Poland\\e-mail: zhab@ift.uni.wroc.pl}
\date{}
\maketitle
\begin{abstract}
Quantum mechanical interference of wave functions leads to some
difficulties if a probability density is
considered as a source of gravity.
We show that an introduction of a quantum
energy-momentum tensor
as a source term in Einstein equations can be consistent with general relativity
if the gravitational waves are quantized.

\end{abstract}

\section{Introduction}
The conceptual difficulties with an introduction of quantum effects
into Einstein gravity   are well-known. It would be a pessimistic attitude
 to  think that  the inconsistencies can be avoided only in a
  complete quantum theory of
gravity and matter .
It has been shown that some semiclassical methods to combine classical gravity
and quantum mechanics contradict experiments and intuition
\cite{page}\cite{ford}. Page and Geilker \cite{page} suggested
that decoherence could help to solve these difficulties.
An interaction with the environment as a source of the decoherence
has been discussed in \cite{cald}-\cite{zurek}.
 In ref.\cite{diosi} Einstein gravity has been discussed and
  a phenomenological stochastic coupling has been introduced in order
 to achieve the decoherence.

      Our aim in this paper is to discuss
      quantum mechanics combined with the linearized quantum gravity
      as a consistent approximation to a hypothetical complete
      quantum theory.
 First, let us outline  a scenario
 implying a physical relevance of  the quantized gravity.
 We consider a model of a gravitational field interacting
 with the quantum complex scalar field. If these fields result from a quantization
  of the   classical field theory then they should satisfy the
  operator equations of motion ($\kappa=8\pi G$, where $G$ is the Newton
  constant)
 \begin{equation}
 \Box_{g}\psi=M^{2}c^{2}\psi
 \end{equation}
 \begin{equation}
 \begin{array}{l}
 R_{\mu\nu}-\frac{1}{2}g_{\mu\nu}R=-\kappa/c^{4} \Big(\partial_{\mu}\overline{\psi}
 \partial_{\nu}\psi   +\partial_{\nu}\overline{\psi}
 \partial_{\mu}\psi
 \cr
  +\frac{1}{2}g_{\mu\nu} g^{\alpha\beta}(\partial_{\alpha}
 \overline{\psi}
 \partial_{\beta}\psi   +\partial_{\beta}\overline{\psi}
 \partial_{\alpha}\psi)-\frac{1}{2}g_{\mu\nu}M^{2}c^{2}\psi\overline{\psi}\Big)
 \end{array}
 \end{equation}
where $g_{\mu\nu}$ is the Riemannian metric, $R_{\mu\nu}$ is
the Ricci tensor constructed from the Christoffel symbols
$\Gamma^{\rho}_{\mu\nu} $ and the wave operator $\Box_{g}$ is
\begin{equation}
\Box_{g}=g^{\mu\nu}
(\partial_{\mu}+g^{\alpha\beta}
\Gamma_{\mu\alpha\beta})\partial_{\nu}
\end{equation}
 Subsequently we can consider a reduction of the quantum field theory
 of the scalar field to
 quantum mechanics . For this purpose we restrict the field equations to a
  fixed $N$-particle
 sector.    Let $\psi_{cl}$ be a solution of the Klein-Gordon
 equation. By means of the asymptotic fields   we can define   the
 one particle state $\vert\psi_{cl}\rangle$ . Then, for the matrix elements
 $\langle 0\vert \psi(x)\vert \psi_{cl}\rangle$ we should obtain  from eq.(1)
 approximately
the Klein-Gordon equation in the metric $g$ .
Taking the expectation value of both sides
of eq.(2) in the state $\vert \psi_{cl}\rangle$ we
obtain the (quantum) Einstein equations determined
by the energy--momentum tensor $\langle \psi_{cl}\vert T_{\mu\nu}(\psi)
\vert \psi_{cl}\rangle $ of the quantum
field theory.

At this stage it is also useful to see the non-relativistic limit of eqs.(1)-(2).
The non-relativistic limit is simple if
 $g^{k0}=0$ and $g^{00}$ is a slowly varying function
 of time. Then, we write
\begin{equation}
g^{00}=-1+V/c^{2}
\end{equation}
and
\begin{displaymath}
 \psi=\exp(-\frac{iM}{\hbar}x_{0})\tilde{\psi}
 \end{displaymath}
As a result we obtain an equation for $\tilde{\psi }$
( where $\triangle_{{\cal M}}^{(3)}$ is the threedimensional
Laplacian on a Riemannian manifold ${\cal M}$)
\begin{equation}
\begin{array}{l}
i\hbar\partial_{t}\tilde{\psi}=
(-\frac{\hbar^{2}}{2M}\triangle^{(3)}_{{\cal M}}-V)\tilde{\psi}
\end{array}
 \end{equation}
i.e., the conventional Schr\"odinger equation in the gravitational field.
Then,  in the limit $c\rightarrow \infty$ we obtain an equation
for $V$
 \begin{displaymath}
( \triangle^{(3)}_{{\cal M}} -\frac{1}{c^{2}}
\partial_{t}^{2}) V=\kappa\vert \tilde{\psi}\vert^{2}
 \end{displaymath}

           We consider  a semiclassical approximation to
  the solution of the Klein-Gordon equation on
 a manifold. The semiclassical
 approximation is determined by the Hamiltonian
 \begin{equation}
 {\cal H}=\frac{1}{2}g^{\mu\nu}(q)p_{\mu}p_{\nu}
 \end{equation}
 A solution of  the Hamiltonian equations of motion determines a
 solution of the Klein-Gordon equation in the leading order of $\hbar$  .
 The Hamilton-Jacobi function
 \begin{displaymath}
 W(x)=\int_{x}^{y} p_{\mu}dq^{\mu}
 \end{displaymath}
  is computed on the trajectory starting from $x$ and ending in $y$.
 We have    in the semiclassical approximation
\begin{equation}
 \psi(x)=\exp(\frac{i}{\hbar}\int_{x}^{y} p_{\mu}dq^{\mu})
 \Big(\det(\frac{\partial q_{\tau}}{\partial x})\Big)^{\frac{1}{2}}
 \phi(q_{\tau}(x))
 \end{equation}
 where $q_{\tau}$ is the trajectory starting from $x$ at $\tau=0$
 and ending in $y$ at $\tau$. We have to exchange $\tau$ in favor of $x_{0}$
 in eq.(7) .

The Hamiltonian equations have a constant of motion
which we express in the form
 \begin{displaymath}
 {\cal H}=\frac{P^{2}}{2M}
 \end{displaymath}
For an approximate solution of the Hamiltonian equations
it will be useful to introduce the frames  $e$
\begin{equation}
e^{\mu}_{a}\eta^{ab}e^{\nu}_{b}=g^{\mu\nu}
\end{equation}
where $\eta=(1,1,1,-1)$ is the Minkowski metric.
We express $e $  by  $\alpha$
\begin{displaymath}
e_{\mu a}=\eta_{\mu c}(\exp \alpha)^{c}_{a}
\end{displaymath}
Then, in the lowest order
\begin{equation}
e_{\mu a}=\eta_{\mu a}+\eta_{\mu c}\alpha^{c}_{a}
\end{equation}
Geodesic equations (which are equivalent to Hamiltonian equations)
can  be expressed in the form
\begin{equation}
d\Big(e_{\mu a}\frac{dq^{\mu}}{d\tau}\Big)=0
\end{equation}
\begin{equation}
d e_{\mu a}+\Gamma^{\nu}_{\mu\alpha}e_{\nu a}dq^{\alpha}=0
\end{equation}
We solve these equations by iteration . So,
 we have  from eq.(10)
 \begin{equation}
\frac{dq^{\mu}}{d\tau}=-e^{\mu a}P_{a}/M
 \end{equation}
 In the zeroth order in $\alpha$ the trajectory
   is a straight line $q=x-P\tau/M$ . We obtain
    a relation between $\tau$ and $x_{0}$ which can be
   applied  in order to
   eliminate $\tau$ from $W$  in eq.(7)
   \begin{displaymath}
   x_{0}-y_{0}=\tau P_{0}/M
   \end{displaymath}

\section{Superposition principle and  gravitational fluctuations}

 We suggest that a quantization of gravitational waves may help to reconcile classical
 Einstein gravity with quantum mechanics. The common way to
 introduce a quantum matter into the classical gravity \cite{ford} is a replacement
 of $T_{\mu\nu}$ by $\langle\psi\vert T_{\mu\nu}\vert\psi\rangle$.
 However, we encounter the Schr\"odinger cat problem with
 such an extension of classical gravity. We
  obtain a superposition of states with a certain density
  determined by $\vert\psi\vert^{2}$. In the classical gravity the metric
  is determined by the energy distribution (possibly random)
  , but  not by the probability density of an interference of
   states of different energies \cite{page}.

We consider a perturbation expansion of eqs.(1)-(2) $g=g^{(cl)}+g^{(q)}$
 with the
approximation of a weak classical field $g^{cl}$. Then, $\Box g^{(q)}=0$.
Hence, the quantum gravitons are approximately moving as waves in
a flat background. Then, expanding in $g^{(cl)}$
\begin{equation}
\Box_{g^{(q)}}g_{00}^{(cl)}-2R_{0}^{(q)\nu} g_{0\nu}^{(cl)}
+2 R^{(q)}_{0\mu\nu 0}g^{(cl)\mu\nu}=\frac{\kappa}{c^{2}}\vert
\psi_{1}(t) +\psi_{2}(t)\vert^{2}
\end{equation}
where we take a superposition  $\psi_{1}+\psi_{2}$ of solutions
of eq.(1). The observed value of the gravitational field is
an average over the quantum fluctuations of the metric
\begin{equation}
\langle\Box_{g^{(q)}}g_{00}^{(cl)}-2R_{0}^{(q)\nu} g_{0\nu}^{(cl)}
+2 R^{(q)}_{0\mu\nu 0}g^{(cl)\mu\nu}\rangle=\frac{\kappa}{c^{2}}
\langle\vert
\psi_{1}(t) +\psi_{2}(t)\vert^{2} \rangle
\end{equation}
In the linear approximation the l.h.s. of eq.(14) is $\Box g^{(cl)}$.
In the semiclassical
approximation for the solution of the
Schr\"odinger equation we obtain a superposition of solutions of eq.(7)
 \begin{displaymath}
 \psi_{1}+\psi_{2}\rightarrow \exp(\frac{i}{\hbar}W^{(1)}(t))\phi(q^{(1)}(t) )
 + \exp(\frac{i}{\hbar}W^{(2)}(t))\phi(q^{(2)}(t))
 \end{displaymath}
 where $q^{(k)}(t)$ are the classical trajectories.
 At the point $x$ the probability density
 $\langle\vert \psi_{t}( x)\vert^{2}\rangle $ is
 equal to the diagonal part of the density matrix
 \begin{equation}
 \rho_{t}=\langle\vert \psi_{t}  \rangle  \langle\psi_{t}\vert \rangle
 \end{equation}

 We assume that the quantum gravity consists of asymptotic fields
   $\alpha^{b}_{a}$
 describing gravitational radiation and eventually some other degrees of freedom which do not
 have asymptotic fields.  We quantize at the moment only the gravitational waves.
In the linear approximation the metric as well as the tetrad $ e$
($\alpha$ in the linear approximation)
satisfy the wave equation
\begin{displaymath}
\Box \alpha^{c}_{a}=0
\end{displaymath}
We quantize the gravitational waves expanding $\alpha$ in the momentum space
\begin{equation}
\begin{array}{l}
\alpha^{b}_{a}(x)=\sqrt{4\pi c}\sqrt{\hbar\kappa/c^{4}}(2\pi)^{-\frac{3}{2}}\int d{\bf k}\vert {\bf k}\vert^{-\frac{1}{2}}
\exp(i{\bf kx})
\cr
\sum_{\zeta=1,2}\Big({\cal E}^{b}_{a}(\zeta,k)C(\zeta,{\bf k})
\exp(-i\vert{\bf k}\vert x_{0} ) +
\overline{{\cal E}^{b}_{a}}(\zeta,k)C(\zeta,{\bf k})^{+}
\exp(i\vert{\bf k}\vert x_{0} ) \Big)
\end{array}
\end{equation}
where
\begin{displaymath}
 [C(\zeta,{\bf k}), C(\zeta^{\prime},{\bf k}^{\prime})^{+}]=
 \delta_{\zeta \zeta^{\prime}}\delta({\bf k}-{\bf k}^{\prime})
 \end{displaymath}
are the creation and annihilation operators.

The Hamiltonian is
\begin{displaymath}
H_{R}=\int d{\bf k}\sum_{\zeta}c\vert{\bf k}\vert C(\zeta,{\bf k})^{+}C(\zeta,{\bf k})
\end{displaymath}
If the gravitational radiation is in equilibrium with light and
matter then it should be described by the Gibbs distribution
\begin{displaymath}
\hat{\rho}_{\beta}=Z^{-1}\exp(-\beta H_{R})
\end{displaymath}
where $\frac{1}{\beta}=KT$ , $K $ is the Boltzmann constant
and $T$ denotes the temperature.
In cosmological models the Gibbs distribution is believed to
be correct  when applied to the primordial
gravitons (present at the earliest stages of the big bang)
which had time to reach an equilibrium with
other particles \cite{weinberg}. We suggest
that these relict gravitons now reach the Earth
and impose the classical behavior of large quantum systems.
During an expansion of the Universe
some gravitons are continuously being created as a result
of a time-dependent gravitational field.
These gravitons will not reach any equilibrium with the primordial
ones. They  have another  energy distribution.
The probability of graviton production is large at small
wave number ${\bf k}$.
Hence, the Planck distribution may be modified
for small wave numbers ${\bf  k}$
 \cite{grischchuk}. For these reasons we
 consider a more general
density matrix
$\hat{\rho}(H_{R}) $ as a function of the graviton energy
$H_{R}$.
 We introduce a parameter $1/b$ (with the dimension of the inverse
 of the energy )  as an energy cutoff .
We could represent $\hat{\rho  }$ by a Fourier-Laplace transform
of the Gibbs distribution $\hat{\rho}_{\mu}$ ($\mu$
may be complex)
\begin{equation}
\hat{\rho}=\int d\mu\gamma(\mu)\hat{\rho}_{\mu}
\end{equation}
Then, for a computation of expectation values we
can use the  methods applied for the Gibbs state ,
e.g.,  by a direct computation through an expansion
in the number states  we find in the Gibbs state
\begin{equation}
f_{\beta}^{PL}(\hbar c\vert{\bf k}\vert)\equiv\langle C^{+}({\bf k})C({\bf k})\rangle_{\beta}=
\Big(\exp(\beta c\hbar|{\bf k}|)-1\Big)^{-1}
\end{equation}
We can see that effectively $1/\beta   $ plays the role of the energy cutoff
in the Planck distribution because $f^{PL}_{\beta}(\hbar kc)$ decays exponentially
fast when   $kc\hbar >1/\beta$. We shall often identify $b$ with $\beta$
in our discussion.

The correlation functions  can  be computed using  eqs.(17) and (18)
\begin{equation}
\begin{array}{l}
Tr\bigg(\alpha^{a}_{c}(t,{\bf x})\alpha^{a^{\prime}}_{c^{\prime}}
(0,{\bf x}^{\prime})\hat{\rho}\bigg)\equiv
G_{\beta}({\bf x},{\bf x}^{\prime};t)^{a a^{\prime}}_{cc^{\prime}}=
\cr
\frac{\hbar \kappa}{2\pi^{2} c^{3}}\int d{{\bf k}}
\delta^{aa^{\prime}}_{cc^{\prime}}\frac{1}{|{\bf k}|}
cos({\bf k}({\bf x}-{\bf x}^{\prime}) )
\cr
\bigg((\frac{1}{2}+f_{b}( \hbar c|{\bf k}|))
cos(c|{\bf k}| t) -\frac{i}{2}sin(c|{\bf k}| t)\bigg)
\end{array}
\end{equation}
where we denoted (a representation of the tensor $\delta$
 and its properties are discussed in \cite{wein})
 \begin{displaymath}
 \delta_{bd}^{ac}(k)= \sum_{\zeta}\overline{{\cal E}^{a}_{b} }(k,\zeta)
 {\cal E}^{c}_{d}(k,\zeta)
 \end{displaymath}
 In eq.(19) $f_{b}$ is the graviton distribution.    Our results for small time and space
separations do not depend essentially on the form of $f_{b}$ if
\begin{displaymath}
f_{b}(k)=\tilde{f}(b k)
\end{displaymath}
and if $\tilde{f}$ decays sufficiently fast, e.g.,
$\vert \tilde{f}(k)\vert\leq Ak^{-6}$ for a large $k$.
For a large time and large space separations the
results depend on the singularity of $\tilde{f}(k)$
at $k=0$  (for the Planck distribution $\tilde{f}\approx k^{-1}$).
The distributions derived in inflationary models \cite{allen}\cite{abbot}
 behave powerlike in some intervals ,e.g.   ,
\begin{displaymath}
\tilde{f}(k)=0
\end{displaymath}
 if $k\geq 1$ and
\begin{equation}
  \tilde{f}(k)=k^{-\sigma}
  \end{equation}
  if $0\leq k \leq 1$ . Such a distribution leads to similar conclusions as
   the Planck distribution.
   However, an introduction of the infrared cutoff $\tilde{f}(k)=0$
  if $k\geq \epsilon$  would
  destroy the decoherence at a sufficiently large  time.

An expectation value in the ground state $\chi$ of the free
gravitational field
 is a special case of eq.(19) corresponding to the limit $\beta
\rightarrow \infty $ (under the assumption $f_{\infty}(k)=0$)
\begin{equation}
\begin{array}{l}
G_{\infty}({\bf x},t;{\bf x}^{\prime},0)
\equiv
<\chi|\alpha (t,{\bf x})\alpha(0,{\bf x}^{\prime})|\chi>=
\cr
\kappa c^{-4}\frac{\hbar c}{4\pi^{2}}\int d{\bf k}\frac{1}{|{\bf k}|}
\cos({\bf k}({\bf x}-{\bf x}^{\prime}))\exp(-ic|{\bf k}| t)
\end{array}
\end{equation}
Comparing eqs.(19) and (21) it can be seen that the first term on the r.h.s.
 of eq.(19) describes the zero point density
(vacuum fluctuations) whereas the second one comes from the thermal gravitons
in equilibrium with the environment. In general, the vacuum fluctuations
cannot be neglected. After a renormalization they contribute
to measurable effects. However, it can be shown \cite{haba}
that renormalized vacuum fluctuations
give a negligible contribution to the decoherence. They are
small in comparison to the black body radiation at moderate temperatures.
Moreover, the renormalized vacuum fluctuation part decreases to zero when the time
becomes large.
We subtract the vacuum fluctuations in eq.(19). After this
subtraction the correlation function becomes real. We can associate
a real random field $\alpha$ with such a correlation function
(the field will be Gaussian in a linear approximation to gravity)
\begin{displaymath}
\langle \alpha^{a}_{b}(x)\alpha^{c}_{d}(x^{\prime})\rangle=G^{ac}_{bd}(x-x^{\prime})
\end{displaymath}
where
\begin{equation}
\begin{array}{l}
 G^{ac}_{bd}(x-x^{\prime})=\hbar\kappa c^{-3}(2\pi)^{-2}\int d{\bf k}
 \vert{\bf k}\vert^{-1}
\delta^{ac}_{bd} (k)
  \cos({\bf k}({\bf x}-{\bf x}^{\prime}))
  \cos((x_{0}-x_{0}^{\prime})\vert{\bf k}\vert) f_{b}
  (c\vert {\bf k}\vert\hbar)
 \end{array}
 \end{equation}
In spherical coordinates $d{\bf k}=2\pi d\theta \sin \theta dk k^{2}$  .
Hence, $G_{th}$ can be expressed in the form
( we skip the tensor $\delta$)
\begin{displaymath}
\begin{array}{l}
G_{th}
 =2\hbar \kappa c^{-3}\vert {\bf x}-{\bf x}^{\prime}\vert^{-1}
 \pi^{-1}
\int_{0}^{\infty} dk sin(k|{\bf x}-{\bf x}^{\prime}|)
  \cos(ckt)  f_{b}(\hbar c k)
   \end{array}
\end{displaymath}

\section{Decoherent effect of gravitons}

It is natural to
associate   gravitons with the decoherence. Gravitons
interact with all particles.
Hence, their decoherence effect would be universal.
We consider the Einstein gravity for weak fields .
We define the partial density matrix (averaged over the gravitons)
 \begin{equation}
\rho_{t}({\bf x},{\bf x}^{\prime})= Tr_{R}\Big(
\langle  {\bf x}\vert \hat{\rho}( H_{R})\exp(-i\frac{t}{\hbar}H_{R})\vert\psi_{t}
\rangle\langle \psi_{t} \vert
 \exp(i\frac{t}{\hbar}H_{R})\vert {\bf x}^{\prime}\rangle\Big)
 \end{equation}
 The trace in eq.(23) can be obtained as an
 expectation value over the Gaussian random field $\alpha$
 (or calculated in the operator formalism by means of  the time-ordered
  products in the Fock space)
\begin{displaymath}
\langle \exp i\alpha J\rangle =\exp( -\frac{1}{2}JGJ)
\end{displaymath}
where the Green functions $G$ depend on the state
under consideration. In particular, in the thermal state
with subtracted vacuum fluctuations $G\rightarrow
G_{th}$ .
  If the vacuum fluctuations were to be taken into account
  then we would need to make the replacement $G_{th}\rightarrow
  G_{\beta}=G_{th}+G_{\infty}$
  and subsequently $G_{\infty}\rightarrow G_{F}=i\triangle_{F}$
  (in the notation of Bjorken and Drell\cite{bjor}). However, the part $
  \int G_{F}d{\bf y}
  d{\bf y}$ contains infinities when the paths intersect .
  After a renormalization the remaining expression gives a negligible
  contribution to the decoherence \cite{haba}.

   We consider an initial state
  \begin{equation}
  \psi({\bf x})=\exp(i{\bf P}^{(1)}{\bf x}/\hbar)\phi ({\bf x})+
\exp(i{\bf P}^{(2)}{\bf x}/\hbar)\phi ({\bf x})
 \end{equation}
 We solve the geodesic equation (12) with the initial position $x$
 and the initial momenta $P^{(k)}$ by iteration. Then, till the
 first order in $\alpha $
\begin{displaymath}
   y^{(k)}_{s}= x-\frac{s}{M}P^{(k)}-\frac{1}{M}\int_{0}^{s}
   \alpha(x-  \frac{\tau}{M}P^{(k)})P^{(k)} d\tau
  \end{displaymath}
  The Green's function on the trajectory (till the zeroth order in $\alpha$)
  is
\begin{displaymath}
\langle\alpha\left(y\left(\tau, x\right)\right)
\alpha\left(y\left(s, x^{\prime}\right)\right)
\rangle=
G({\bf x}-{\bf x}^{\prime}-
\frac{\tau-s}{M}{\bf P},x_{0}-x_{0}^{\prime}-\frac{\tau-s}{M}P_{0})
\end{displaymath}
    In our approximation the time evolution is
\begin{displaymath}
\begin{array}{l}
  \psi({\bf x})\rightarrow\psi_{t}({\bf x})=
  \exp(i P^{(1)} y^{(1)}_{t}/\hbar)
\phi ({\bf y}^{(1)}_{t})
 +\exp(i P^{(2)} y^{(2)}_{t}/\hbar)
\phi ({\bf y}^{(2)}_{t})
\end{array}
 \end{displaymath}
 where  $y_{t}^{(k)}(x)=({\bf y}_{t}^{(k)}({\bf x}),ct) $ and
 \begin{displaymath}
 P^{(k)}=\Big({\bf P}^{(k)},\sqrt{M^{2}c^{2}+({\bf P}^{(k)} )^{2}}\Big)
 \end{displaymath}
Then, a calculation of the trace (23) leads to the formula
( we assume that $\phi$ is a slowly varying function, hence
in its argument it is sufficient to calculate  $y^{(k)}$ in the
lowest order in $\alpha$)
 \begin{equation}
  \begin{array}{l}
  \langle\mid\psi_{t}({\bf x})\mid^{2}\rangle=
  \vert\phi({\bf x}-\frac{t}{M}{\bf P}^{(1)})\vert^{2}
+  \vert\phi({\bf x}-\frac{t}{M}{\bf P}^{(2)})\vert^{2}     +
\cr
\Big(\overline{\phi({\bf x}-\frac{t}{M}{\bf P}^{(2)})}
   \phi({\bf x}-\frac{t}{M}{\bf P}^{(1)})
  \exp\Big(\frac{i}{\hbar}( P^{(2)}- P^{(1)}) x\Big)
\cr
+\overline{\phi({\bf x}-\frac{t}{M}{\bf P}^{(1)})}
   \phi({\bf x}-\frac{t}{M}{\bf P}^{(2)})
  \exp\Big(-\frac{i}{\hbar}( P^{(2)}- P^{(1)}) )x\Big)
  \cr
  \exp\Big(\frac{1}{M^{2}\hbar^{2}}\int_{0}^{t}
{\bf P}^{(1)}{\bf P}^{(1)}  G_{th}(\frac{s}{M}{\bf P}^{(1)}-\frac{\tau}{M}{\bf P}^{(2)}
,P_{0}^{(1)}s/M-P_{0}^{(2)}\tau/M)
{\bf P}^{(2)}{\bf P}^{(2)}dsd\tau
\cr
-\frac{1}{2M^{2}\hbar^{2}}\int_{0}^{t}
  {\bf P}^{(1)}{\bf P}^{(1)}G_{th}(\frac{s}{M}{\bf P}^{(1)}
  -\frac{\tau}{M}{\bf P}^{(1)},P_{0}^{(1)}(s-\tau)/M)
  {\bf P}^{(1)} {\bf P}^{(1)}dsd\tau
  \cr
  -\frac{1}{2M^{2}\hbar^{2}}\int_{0}^{t}
  {\bf P}^{(2)}{\bf P}^{(2)}G_{th}(\frac{s}{M}{\bf P}^{(2)}-\frac{\tau}{M}{\bf P}^{(2)}
  ,P_{0}^{(2)}(s-\tau)/M) {\bf P}^{(2)}  {\bf P}^{(2)}
  d{\tau}ds \Big)
  \cr
  \equiv
  \rho_{t}^{(1)}+\rho_{t}^{(2)} +\rho_{t}^{(12)}
\end{array}
  \end{equation}
  The notation ${\bf P}{\bf P}$ is only symbolic, it means
  that we must sum  the indices of ${\bf P}$ with the indices
  of the Green's function $G_{th}$.

  If the off-diagonal terms vanish then the density of a superposition
  of wave functions is a sum of the densities.
Classically the packets move apart and each of them is independently
a source of a  gravitational field. Without the decoherence
there would be a superposition of the states of the packets.

  It is easy to estimate the behavior of eq.(25) for a small time
  and either ${\bf P}^{(1)}\parallel {\bf P}^{(2)}$ or
  ${\bf P}^{(1)}\perp {\bf P}^{(2)}$.
  Let us denote in  eq.(25)
  \begin{displaymath}
  \vert\rho^{(12)}\vert=2\exp(-S_{12}(P))\vert  \phi({\bf x}-\frac{t}{M}{\bf P}^{(1)})
   \phi({\bf x}-\frac{t}{M}{\bf P}^{(2)})\vert
  \end{displaymath}
  where
\begin{displaymath}
S(P)=P^{a}P_{b}S_{ad}^{bc}P_{c}P^{d}
\end{displaymath}
 Then, for a small time we can set $s=\tau=0$. There remains
  (under the assumption $f_{b}(u)=\tilde{f}(b u)$ )
  \begin{equation}
  \begin{array}{l}
  S_{12}({\bf P})=t^{2}M^{-2}\hbar^{-2}\vert ({\bf P}^{(1)})^{2}-
  ( {\bf P}^{(2)})^{2}\vert^{2}G_{th}(0,0)
  \cr
  =At^{2}M^{-2}\hbar^{-3}\vert ({\bf P}^{(1)})^{2}-  ( {\bf P}^{(2)})^{2}\vert^{2}
   \kappa c^{-5}\beta^{-2}\int_{0}^{\infty}du uf_{b}(u/b)
   \cr
   =\tilde{A}\vert ({\bf P}^{(1)})^{2}-   ({\bf P}^{(2)})^{2}\vert
(\frac{t}{M})^{2}l_{dB}^{-2}
\vert({\bf P}^{(1)})^{2}-   ({\bf P}^{(2)})^{2}\vert \hbar^{-2}
 L_{PL}^{2} \end{array}
 \end{equation}  
In eq.(26) $A$ and $\tilde{A}$ are constants of order 1, $\l_{dB}=\hbar c b=
l_{C} bmc^{2}$ will be called de Broglie
length because if $\beta=b$ then $l_{dB}=\hbar c\beta$
is the wave length of a particle in a medium at temperature T, $l_{C}=\frac{\hbar}{mc}$ is the Compton length
and $m$ is the electron mass as our basic mass unit. Then, $\frac{\hbar}{\vert {\bf P}\vert}$
is particle's wave length at the momentum ${\bf P}$
, $L_{PL}=\sqrt{\hbar\kappa/c^{3}}$ is the Planck length.

 We can see that the decoherence is determined by the relation
 of the particle's wave length and the length of particle's trajectory
 to the Planck's length and de Broglie's length.
For a large time the calculations are more involved.
 Let us denote by $S_{12}$ the expression $S_{12}(P)$ without
 the fourlinear momenta. Then, we obtain a rather complicated formula
 for $S_{12}$ (we omit the tensors $\delta$)
 \begin{equation}
 \begin{array}{l}
 S_{12}=\frac{\kappa}{2M^{2}c^{4}\hbar^{2}} \frac{\hbar c}{2\pi^{2}}
\int d{\bf k} k^{-1} \tilde{f}\left(b\hbar c
  k\right)
 \cr
 \Big( \left({\bf P}^{(2)}\right)^{2}\left({\bf k} {\bf P}^{(2)}/M+
  ck\right)^{-2}\left(1-\cos\left(t{\bf k} {\bf P}^{(2)}/M+
 t ck \right)\right)
 \cr
  +\left({\bf P}^{(2)}\right)^{2}\left({\bf k} {\bf P}^{(2)}/M-
  ck\right)^{-2}\left(1-\cos\left(t{\bf k} {\bf P}^{(2)}/M-
 t ck \right)\right)
 \cr
+\left({\bf P}^{(1)}\right)^{2}\left({\bf k} {\bf P}^{(1)}/M+
  ck\right)^{-2}\left(1-\cos\left(t{\bf k} {\bf P}^{(1)}/M+
 t ck \right)\right)
 \cr                     
  +\left({\bf P}^{(1)}\right)^{2}\left({\bf k} {\bf P}^{(1)}/M-
  ck\right)^{-2}\left(1-\cos\left(t{\bf k} {\bf P}^{(1)}/M-
 t ck \right)\right)
 \cr
 +2({\bf P}^{(1)}{\bf P}^{(2)})({\bf P}^{(1)}{\bf k})({\bf P}^{(2)}{\bf k})/M^{2}
 \left(c^{2}k^{2}-\left({\bf P}^{(1)}{\bf k}\right)^{2}/M^{2}\right)^{-1}
 \left(c^{2}k^{2}-\left({\bf P}^{(2)}{\bf k}\right)^{2}/M^{2}\right)^{-1}
  \cr
  \left(1-\cos\left(t{\bf k}\left( {\bf P}^{(1)}- {\bf P}^{(2)}\right)/M
  \right)\right)
  \cr
  -({\bf P}^{(1)})^{2}({\bf P}^{(2)})^{2}  (ck+{\bf P}^{(1)}{\bf k}/M)^{-1}
     (ck+{\bf P}^{(2)}{\bf k}/M)^{-1} (1-\cos(ckt+t{\bf P}^{(1)}{\bf k}/M))
     \cr
  -({\bf P}^{(1)})^{2}({\bf P}^{(2)})^{2}  (ck-{\bf P}^{(1)}{\bf k}/M)^{-1}
     (ck-{\bf P}^{(2)}{\bf k}/M)^{-1} (1-\cos(ckt-t{\bf P}^{(1)}{\bf k}/M))
\cr
-({\bf P}^{(1)})^{2}({\bf P}^{(2)})^{2}  (ck+{\bf P}^{(1)}{\bf k}/M)^{-1}
     (ck+{\bf P}^{(2)}{\bf k}/m)^{-1} (1-\cos(ckt+t{\bf P}^{(2)}{\bf k}/M))
 \cr
 -({\bf P}^{(1)})^{2}({\bf P}^{(2)})^{2}  (ck-{\bf P}^{(1)}{\bf k}/M)^{-1}
     (ck-{\bf P}^{(2)}{\bf k}/M)^{-1} (1-\cos(ckt-t{\bf P}^{(2)}{\bf k}/M))
 \Big)
 \end{array}
 \end{equation}
 It can be seen from eq.(27) that if $\tilde{f}(0)=0$
 then owing to the Lebesgue theorem $
 S_{12} (t)$ is bounded in $t$. If $\tilde{f}(k)$ is singular at $k=0$
 then $S_{12}(t)$ grows for a large $t$.
In order to investigate this case in
more detail let us consider first the Planck distribution $f_{\beta}^{PL}$ .
We neglect
 ${\bf P}/Mc$ in the denominator of eq.(27) and consider
  either ${\bf P}^{(1)}\parallel {\bf P}^{(2)}$ or
  ${\bf P}^{(1)}\perp {\bf P}^{(2)}$.
   Then, we obtain

\begin{equation}
  \begin{array}{l}
  S_{12}(P)=\frac{\kappa}{M^{2}c^{4}\hbar^{2}}\frac{\hbar }{c\pi}
    \vert ({\bf P}^{(2)})^{2}-({\bf P}^{(1)})^{2}\vert  ^{2}
  \int_{0}^{\infty}\frac{dk}{k}
  (\exp(\beta\hbar ck)-1)^{-1}
  \left(1-\cos\left(tck \right)\right)
  \cr
  =\frac{\kappa}{M^{2}c^{4}\hbar^{2}}\frac{\hbar }{c\pi}
    \vert ({\bf P}^{(2)})^{2}-({\bf P}^{(1)})^{2}\vert  ^{2}ct
    \int_{0}^{1}d\alpha\int_{0}^{\infty}dk
       (\exp(\beta\hbar ck)-1)^{-1} \sin(\alpha ckt)
       \cr
   =\frac{\kappa}{M^{2}c^{4}\hbar^{2}}\frac{\hbar t}{\pi}
    \vert ({\bf P}^{(2)})^{2}-({\bf P}^{(1)})^{2}\vert  ^{2}
    \int_{0}^{1}d\alpha
       (\frac{\pi}{2\beta\hbar c}\coth(\frac{\pi\alpha t}{\beta\hbar})-
       \frac{1}{2\alpha c})
       \cr
   
   =\frac{\kappa}{M^{2}c^{4}\hbar^{2}}\frac{\hbar }{2c\pi}
    \vert ({\bf P}^{(2)})^{2}-({\bf P}^{(1)})^{2}\vert  ^{2}
    \ln\Big(\frac{\beta\hbar}{\pi t}\sinh
       (\frac{t\pi}{\beta\hbar })\Big)
       \cr
       \approx                                   
\frac{ct }{2l_{dB}} L_{PL}^{2}\hbar^{-2}
    \vert ({\bf P}^{(2)})^{2}-({\bf P}^{(1)})^{2}\vert  ^{2}M^{-2}c^{-2}
    \end{array}
    \end{equation}
    for a large t.
    Here, the formula
     3.911 of Gradshtein and Ryzhik \cite{grad} has been applied
\begin{displaymath}
\int_{0}^{\infty}du \sin(au)\Big(\exp(\beta u )-1\Big)^{-1}=
\frac{\pi}{2\beta}\coth(\frac{\pi a}{\beta})-\frac{1}{2a}
\end{displaymath}
For the inflationary distribution (20) $\frac{ct}{l_{dB}}$
in eq.(28) is replaced
by $(\frac{ct}{l_{dB}})^{\sigma}$ .

As a result of the decoherence the gravitational field
is determined by the density distribution of two separate quantum particles.
Hence, we have obtained a classical addition of probabilities
instead of the quantum addition of amplitudes.
This conclusion follows from the results (25)-(28) which should be inserted into
eq.(14).
 
  Finally, let us discuss   numerical estimates on
  some expressions in this paper.    First, let us assume
  that the graviton distribution is determined by
  the Planck distribution   (as suggested by Weinberg \cite{weinberg}).
   The decoherence time
 depends on the ratio of the Planck length to the de Broglie
wave length (as could have been expected because we have two
universal length scales
at finite temperatures).     To see the meaning of
$l_{dB}$ let us write $KT=mv_{T}^{2}$ then $l_{dB}=l_{C}
\frac{c^{2} }{v_{T}^{2}}$ (the Compton length multiplied by the ratio
of the rest energy to the kinetic energy).
For numerical estimates it is
useful to introduce the Planck temperature $T_{PL}$ determined
by $L_{PL}=l_{dB}$, then  $\beta^{2}c^{5}=\kappa/\hbar$
 and $L_{PL}/l_{dB}=T/T_{PL}$
(note that $T_{PL}=1.3\times 10^{32} Kelvin$).
Instead of the de Broglie length we could use the temperature
independent length scale : the Compton length  $l_{C}$ (choosing
the electron mass $m$ as a mass unit). The Compton length
comes in a natural way because the ratio of the gravitational force
to the Coulomb force between electrons is
\begin{displaymath}
F_{grav}/F_{Coul}=\frac{1}{\frac{e^{2}}{\hbar c}}(L_{PL}/l_{C})^{2}\approx
\frac{1}{\frac{e^{2}}{\hbar c}}10^{-46}
\end{displaymath}
 In particular, we can write for a small time
\begin{displaymath}
\rho_{t}\approx \exp\Big(-\frac{M^{2}}{m^{2}}(\frac{L_{PL}}{l_{C}})^{2}l_{dB}^{-2}
(ct)^{2}(\frac{P}{Mc})^{4}\Big)
\end{displaymath}
 The  exponential is of the order $(10^{-23}\frac{M}{m})^{2}(\frac{v}{c}  )^{4}$.
 We can see that we need   the number of particles
 $N\equiv\frac{M}{m}\approx   10^{23}$ if the decoherence is to be
 visible .

If instead of the Planck distribution we have the (simplified) inflationary
distribution (20) then the behavior of $S_{12}$ is determined
 by the integral (with a certain constant $B$)
\begin{displaymath}
S_{12}=B\int_{0}^{b} dk k^{-\sigma-1}\left(1-\cos\left(kct\right)\right)
\end{displaymath}
Under the      assumption
  $0\leq \sigma<2$
this integral behaves as $(\frac{ct}{l_{dB}})^{\sigma}$ for a large $t$
  and as $(\frac{ct}{l_{dB}})^{2}$ for a small $t$
  (we use the notation $l_{dB}=bc\hbar$).
   In inflationary models
   $\sigma $ is different in various frequency ranges and for the lowest
  frequency range  $\sigma=2$. For such a $\sigma$ the integral
  $S_{12}$ is infrared divergent. At the large
  wave length $1/k$ there can be a sharp cutoff
   or a continuous change of the behavior
   of $f$ leading to the finite integral for $S_{12}$.
  This is a sensitive problem in the spectral theory of
  gravitational waves \cite{zeld}\cite{maia} concerning
  waves with length larger than the actual Hubble horizon length.
  It is interesting that in this way the decoherence is associated
  with the prospective evolution of the Universe.

  \end{document}